\documentclass[12pt]{iopart}
\usepackage[OT4]{fontenc}

\usepackage{graphicx}
\usepackage{epstopdf}
\usepackage{iopams}
  \expandafter\let\csname equation*\endcsname\relax
  \expandafter\let\csname endequation*\endcsname\relax 
\usepackage{amsmath,bm}

\newcommand{\rl}{\rangle\!\langle}

\DeclareMathOperator{\im}{Im}
\DeclareMathOperator{\atan}{atan}
\begin{document}
\title{Electronic and optical properties of non-uniformly shaped InAs/InP quantum dashes}
\author{Piotr Kaczmarkiewicz and Pawe{\l} Machnikowski}
\address{Institute of Physics, Wroc{\l}aw University of Technology, 50-370 Wroc{\l}aw, Poland}
\ead{\mailto{piotr.kaczmarkiewicz@pwr.wroc.pl},\mailto{pawel.machnikowski@pwr.wroc.pl}}

\begin{abstract}
We theoretically study the optical properties and the electronic structure of highly elongated InAs/InP quantum dots (quantum dashes) and show how carrier trapping due to geometrical fluctuations of the confining potential affects the excitonic spectrum of the system. We focus on the study of the optical properties of a single exciton confined in the structure. The dependence of the absorption and emission intensities on the geometrical properties (size and position) of the trapping centre within the quantum dash  is analysed and the dependence of the degree of linear polarization on these geometrical parameters is studied in detail. The role of Coulomb correlations for the optical properties of these structures is clarified.
\end{abstract}
\submitto{\SST}
\maketitle

\section{Introduction}

Quantum dashes (QDashes) are highly elongated quantum dot structures (lateral aspect ratio greater than~4) formed spontaneously in a process of self-assembled growth \cite{reithmaier07} or by means of droplet epitaxy \cite{jo10}. They show emission at wavelengths in the region of the third telecommunication window, broad gain, high degree of tunability and, in some cases, a high surface density \cite{reithmaier07,lelarge07,djie08,dery04,sauerwald05,frechengues99,hein09}. Such properties make QDashes not only favourable over other quasi-zero dimensional structures for present telecommunication applications (e.g., InP structures commonly used in high performance lasers and optical amplifiers operating at 1.55~$\mu$m \cite{reithmaier07,lelarge07,rosales11}) but also promising for possible future single-photon technologies.

Optimal design and manufacturing of QDash based devices requires the understanding of their electronic structure and its relation to their morphology. Structural data \cite{reithmaier07} reveal that the geometrical shape of the real QDashes is non-uniform, with zig-zag bends and cross-section size fluctuations. Both these shape irregularities, as well as compositional inhomogeneities and the related non-uniform strain distribution can induce an additional trapping potential within a QDash structure. Recently, by comparison of experimental study and theoretical modelling, we have shown that the observed polarization properties of a QDash ensemble can by explained only if additional carrier trapping is assumed within the QDash volume \cite{musial12}. Other experimental results also revealed that the properties of single QDashes resemble features characteristic of strong confinement regime \cite{sek09}, which is in contrast to the relatively large volume of these structures. Similar effects of confinement in an effectively low volume, as compared to the whole structure, have also been observed in V-shaped quantum wires \cite{guillet03}. Non-uniformities in quantum wire cross-section have also been subject of studies which shown that surface states can localize on the deformation \cite{cantele00,ortix11,marchi05}.

While both electronic properties as well as optical transitions of uniformly-shaped QDashes \cite{miska04,wei05,planelles09} and ensemble polarization properties of non-uniformly shaped ones \cite{musial12} have been modelled previously, the presence of additional confinement has not been studied systematically. 
The highly elongated, uniform nanostructures have been analyzed using multiband $k\cdot p$ theory \cite{andrzejewski10,saito08} and tight-binding approach \cite{sheng08} which were able to reproduce the observed optical properties of the system. However, the details of the electronic structure of non-uniformly shaped QDashes have yet to be studied. The single QDash properties are expected to differ significantly depending on the strength of the trapping centre providing the additional confinement in the system.

In our previous paper \cite{musial12} we have analysed a minimal set of parameters corresponding only to the experimental setup and studied the ensemble emission from a system consisting of symmetrical ($D_{2}$ symmetry) structures with arbitrary size distribution. Due to such a treatment the individual properties of exciton eigenstates have not been shown. In order to assess the influence of carrier trapping on overall electronic and optical properties a more detailed study, including the symmetry breaking of a structure, is necessary.

In this paper, we analyse in detail the relation between the morphology of an InAs/InP QDash and its optical properties. We study how the transition from a uniform structure with no additional trapping to a structure with a strongly trapped exciton ground state changes the properties of the system. We focus on individual exciton properties and analyse not only the depth of the trapping centre and its influence on system spectrum, but also the effect of QDash symmetry breaking by shifting its position. Additional trapping inside the elongated structure is shown to strongly affect the optical properties both qualitatively and quantitatively. The presence of strongly trapped states not only influences the dipole moments for optical transitions but also the electron-hole Coulomb interaction, leading to nontrivial behaviour of the transition rates and of the degree of linear polarization (DOP).

The paper is organized as follows. In section \ref{s:model}, we provide the model and the theoretical framework of our study. In section \ref{s:results}, we present the results of our theoretical analysis concerning the electronic structure and optical properties of the system. In section \ref{ss:sym}, we focus on a system with a trapping centre placed symmetrically in the structure. We model electronic and optical properties of a single QDash and also show how Coulomb interaction affects the spectrum of the system. Next, in section \ref{ss:asym}, we show how the properties of the QDash are modified if we introduce $D_2$ symmetry breaking by shifting the trapping potential fluctuation from the centre of the structure. We conclude the paper in section \ref{s:conclusions}.

\section{The Model}\label{s:model}
We consider a highly elongated quantum dot-like structure with the length to width ratio greater than 5 and characterized by the width and thickness variation along its length, located either symmetrically in the centre of the structure (preserving $D_2$ symmetry) or shifted from the centre, breaking this additional symmetry. We introduce a QDash thickness variation as the trapping mechanism, as it allows us to easily manipulate the strength of additional confinement in the structure. Such an approach has been successful in qualitatively reproducing the temperature dependence of the degree of linear polarization of an ensemble of InAs/InP quantum dashes \cite{musial12}.
\begin{figure}[tb]
\begin{center}
\includegraphics[width=100mm]{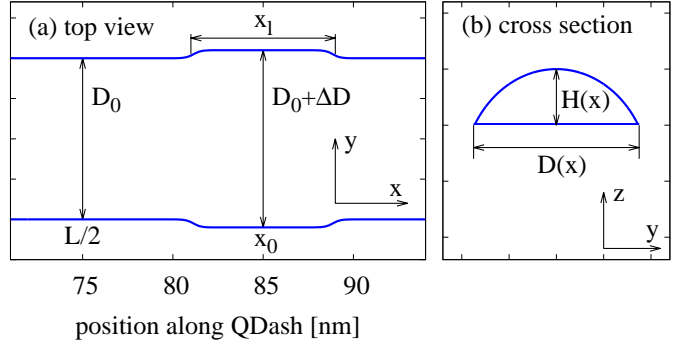}
\end{center}
\caption{\label{fig:schemat} The characteristic features of the QDash geometry: (a) top view of the region of the structure widening, (b) cross section of the structure. The cross section is a circular segment, with a fixed height to base width ratio of $H/D=1:5.5$.}
\end{figure}

We use a single-band effective mass and envelope wave function description. The Hamiltonian of a single carrier is then
\begin{displaymath}
H_{c}=-\frac{\hbar^2}{2m^*_{c}} \Delta + V(\bm{r}),
\end{displaymath}
where $c$ denotes the carrier type (electron or hole), and $m^*_c$ is the effective mass of the carrier in a single band approximation. The QDash confinement potential is modelled as a three dimensional  potential well, described by the $V(\bm{r})$ term, with the shape reproducing the essential features of the realistic QDash geometry. Particularly important is the presence of a width fluctuation (widening) [figure~\ref{fig:schemat}(a)], which provides a trapping centre for the carrier ground state, and effectively increases the confinement. The cross section of the structure is assumed to be a circular segment [figure~\ref{fig:schemat}(b)], with the base width changing along the QDash according to  the model function
\begin{displaymath}
D(x)= D_{0}+\frac{\Delta D (1+4e^{-b})}{1+4e^{-b}\cosh[2b(x-x_0)/x_{\mathrm{l}}]},
\end{displaymath}
where $x$ is the coordinate along the QDash structure,
$D_0$ is the QDash base width away from the widening, $\Delta D$ is the magnitude of the width fluctuation, $x_\mathrm{l}$ is the length of the fluctuation and $x_0$ is the position of the centre of the widening. The $b$ parameter defines the shape of the widening (we choose $b=20$). We define the widening parameter as the ratio of the excess width to the QDash width away from the trapping centre, $\lambda=\Delta D/D_0$.  
%
%
The QDash width to height ration is kept constant, $D = \alpha H$, with $\alpha=5.5$, which is typical for these structures \cite{sauerwald05}. The total length of the structure is set to $150$ nm and the length to width ratio is $6:1$. 

The material parameters used in our calculations are those for InAs/InP structures \cite{lawaetz71}. The effective band offsets (including strain effects) between the materials defining the depth of the confinement potentials, are taken as 400~meV and 250~meV for electrons and holes, respectively, and the effective masses are $0.037 m_e$ for electrons and $0.33 m_e$ for holes.

In order to model single carrier envelope wave functions we use a variational method and follow the adiabatic approximation \cite{musial12,wojs96}, as the confinement along the $x$ direction is much weaker than the confinement in the other directions. First, for a set of positions on a fixed grid along the elongation direction ($x$) we variationally minimize the single-particle Hamiltonian,
\begin{displaymath}
H_{yz}= -\frac{\hbar^2}{2m^*}
\left( \frac{\partial^{2}}{\partial y^{2}}
  +\frac{\partial^{2}}{\partial z^{2}} \right)  + V(\bm{r}),
\end{displaymath}
in the class of two dimensional harmonic oscillator ground state wave functions
\begin{eqnarray}
\lefteqn{\phi_{0}(y,z;x)=}\nonumber \\
&&\frac{1}{\sqrt{l_z(x) l_y(x) \pi }} \exp
\left\{ -\frac{[z-z_0(x)]^2}{2l^{2}_z(x)}
-\frac{y^2}{2l^{2}_y(x)} \right\},\nonumber
\label{wzor:psiZ}
\end{eqnarray}
from which we obtain the set of variational parameters $l_y(x)$, $l_z(x)$ and $z_0(x)$ corresponding to the characteristic confinement lengths for the $y$ and $z$ directions, and the centre of the wave function along the $z$ direction, respectively. In the next step we generate a set of effective potentials along the direction of the elongation,
\begin{equation}\label{wzor:EffPot}
\epsilon_n(x)=  \int dz \int dy
\phi_{n}^{*}(y,z;x)H_{yz}\phi_{n}(y,z;x),\nonumber
\end{equation}
where $\phi_n$ is the wave function of a 2D harmonic oscillator with the same variational parameters as before but representing the $n$-th state along the $y$ direction. 
\begin{figure}[tb]
\begin{center}
\includegraphics[width=100mm]{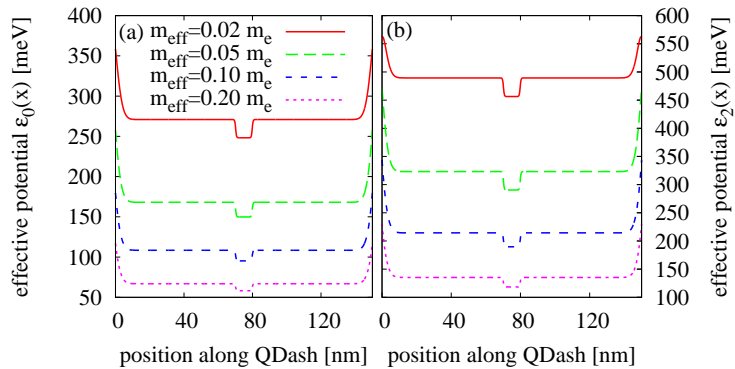}
\end{center}
\caption{\label{fig:potencjal} The calculated effective potentials for different values of carrier effective mass: (a) $\epsilon_{0} (x)$, (b) $\epsilon_{2} (x)$ for the widening parameter $\lambda=0.1$. For lighter carrier the depth of potential fluctuation is higher than that for a heavier carrier (for $m_\mathrm{eff}=0.02$ the potential fluctuation depth is $22$~meV for $\epsilon_{0}(x)$ and $33$~meV for $\epsilon_{2}(x)$, and for $m_\mathrm{eff}=0.2$ --- $9$ and $17$~meV, respectively).}
\end{figure}
A set of effective potentials for different effective masses has been presented in figure \ref{fig:potencjal}(a) for $n=0$ and figure \ref{fig:potencjal}(b) for $n=2$. As we can see for lighter carrier the same widening ($\lambda=0.10$) leads to slightly deeper effective potential fluctuation. Also the depth of the trapping potential fluctuation is deeper for states excited along $y$ direction ($n=2$). The depth of the potential fluctuation can be adjusted with $\lambda$ parameter and grows with growing value of $\lambda$. Next, the obtained effective potentials are used in a one dimensional eigenvalue equations describing the system state in a QDash elongation direction (we suppress the index denoting the carrier type for clarity)
\begin{equation}\label{wzor:RnieEfektywn}
\left [ -\frac{\hbar^{2}}{2m^*}\frac{\partial^{2}}{\partial x^{2}}
+ \epsilon_n(x) \right ] f_{nm}(x) = E_{nm} f_{nm}(x).
\end{equation}
The complete approximate envelope wave functions are then $\psi_{nm}(x,y,z)=\psi_{n}(y,z;x)f_{nm}(x).$

The single carrier envelope wave functions can now be used to construct the excitonic product basis. For simplicity, from now on we use a single index to describe a carrier state characterized by the quantum numbers $n$ and $m$. The Hamiltonian describing a single exciton confined in the structure is
\begin{eqnarray}
H &=& \sum_{i} E_{i}^{(\mathrm{e})} |i_\mathrm{e} \rl i_\mathrm{e}| +\sum_{i}
E_{i}^{(\mathrm{h})} |i_\mathrm{h} \rl i_\mathrm{h}| \nonumber \\
  & &+ \sum_{ijkl} V_{ijkl} |i_\mathrm{e} j_\mathrm{h}\rl k_\mathrm{e} l_\mathrm{h}|,
\label{wzor:ham_ex}
\end{eqnarray}
where $E_i^{(\mathrm{e,h})}$ are the eigenenergies calculated from (\ref{wzor:RnieEfektywn}) and $V_{ijkl}$ are the matrix elements for the electron-hole interaction,
\begin{eqnarray*}
V^{\mathrm{e-h}}_{ijkl} &=& \langle ij | H_{\mathrm{e-h}} | kl \rangle \\
&=& - \int d^3 r_e \int d^3r_h
\psi_{i}^{*(\mathrm{e})} (\bm{r_{\mathrm{e}}})
\psi_{j}^{*(\mathrm{h})} (\bm{r_{\mathrm{h}}})\\
&& \times \frac{e^2}{4 \pi \varepsilon \varepsilon_0}
\frac{1}{|\bm{r_e}-\bm{r_h}|}
\psi_{k}^{(\mathrm{h})}(\bm{r_{\mathrm{h}}})
\psi_{l}^{(\mathrm{e})}(\bm{r_{\mathrm{e}}}),
\end{eqnarray*}
where $\varepsilon_0$ is the vacuum permittivity and $\varepsilon$ is the relative dielectric constant of the QDash material ($\varepsilon = 14.6$ for InAs).

In the calculations of the dipole moments, we assume mainly heavy hole character of the hole states, with only a small admixture of light holes (see Reference \cite{musial12} for details). The components of the interband dipole moment corresponding to the transition to the exciton state $\beta$ for the polarization parallel ($l$) and transverse ($t$) to the direction of the elongation of the structure are
\begin{equation}\label{wzor:dl}
d_{l,t}^{(\beta)} = \mp d_0 \frac{i\pm1}{2}\alpha_{3/2,1/2}^{(\beta)}
+ d_0 \frac{1\mp i}{2\sqrt{3}}\alpha_{-1/2,1/2}^{(\beta)}\nonumber,
\end{equation}
where the upper and lower signs correspond to '$l$' and '$t$', respectively. 
The parameters $\alpha_{-1/2,1/2}$ and $\alpha_{3/2,1/2}$ are the oscillator strengths
for light and heavy hole contributions, respectively, and are defined as
\begin{equation}\label{wzor:defalpha}
\alpha_{3/2,1/2}^{(\beta)} =\sum_{ij} c_{ij}^{(\beta)}
\int d^3r \psi_{i}^{(\mathrm{h})}({\bm{r}})\psi_{j}^{(\mathrm{e})}({\bm{r}})\nonumber
\end{equation}
and
\begin{equation}\label{wzor:defalpha2}
\alpha_{-1/2,1/2}^{(\beta)} =- \frac{1}{\Delta E_{\mathrm{lh}}} \sum_{ij} c_{ij}^{(\beta)}
\int d^3r \psi_{i}^{(\mathrm{h})}({\bm{r}})R_k \psi_{j}^{(\mathrm{e})}({\bm{r}}),\nonumber
\end{equation}
where $c_{ij}^{(\beta)}$ are coefficients for the expansion of the exciton state $\beta$ into the product basis states $\psi_i^{(\mathrm{e})}({\bm{r_e}})\psi_j^{(\mathrm{h})}({\bm{r_h}})$ calculated by diagonalization of the Hamiltonian given in \Eref{wzor:ham_ex},
$R_k$ is the kinetic part of the Kane Hamiltonian matrix element 
coupling the $3/2$ spin heavy-hole subband with the $-1/2$ light hole subband \cite{andrzejewski10} and $\Delta E_{\mathrm{lh}}$ is the average separation between light and heavy hole states. Note that we use the standard definition of the Kane model basis functions \cite{winkler03} consistent with the general theory of the angular momentum \cite{sakurai94}, which are
\begin{eqnarray}
u^{1}_{HH}  & = & -\frac{1}{\sqrt{2}} ( | x \rangle + i | y \rangle ) \chi_{\uparrow} ,\nonumber \\
u^{2}_{HH}  & = & \frac{1}{\sqrt{2}} ( | x \rangle - i | y \rangle ) \chi_{\downarrow}, \nonumber \\
u^{1}_{LH}  & = & \frac{1}{\sqrt{6}} ( - | x \rangle  - i  | y \rangle  )\chi_{\downarrow} + \sqrt{\frac{2}{3}} | z \rangle  \chi_{\uparrow} ,\nonumber \\
u^{2}_{LH}  & = & \frac{1}{\sqrt{6}} ( | x \rangle - i | y \rangle )\chi_{\uparrow} +  \sqrt{\frac{2}{3}} | z \rangle  \chi_{\downarrow}, \nonumber
\end{eqnarray}
where $|x\rangle$, $|y\rangle$, $|z\rangle$ are $p$-like valence band functions and $\chi_{\downarrow,\uparrow}$ are eigenstates of $\sigma_{z}$ Pauli matrix.

The structure morphology is most clearly manifested in the optical emission via its impact on the polarization properties of luminescence \cite{musial12,tonin12}. The light hole admixture corresponds to an additional interband transition dipole which interferes with the heavy hole transition dipole leading to an overall elliptical polarization of the emitted radiation. The polarization ellipse can be characterized by two angles $\chi$ and $\Psi$, determining the ellipticity (with $\tan \chi$ describing the ratio of minor to major axes of the ellipse) and the direction of major axis, respectively \cite{saleh07}. For a sufficiently uniform strain field, the contribution of the strain related term of Kane Hamiltonian matrix element to the degree of linear polarization is expected to be constant \cite{musial12} and will be disregarded. Within the constant strain field approximation, the phase difference between longitudinal and transverse polarizations is constant and equals $\pi/2$, therefore  the direction of the polarization ellipse major axis is parallel to the structure elongation direction. Small differences between those two directions have been observed in the experiment \cite{tonin12} but this effect is beyond the scope of the current paper as it is attributed to strain field anisotropy \cite{ohno11}. In the present case, the $\chi$ parameter for the emission from a single exciton state $\beta$ is simply
$\chi^{(\beta)}= \atan [|d_t^{(\beta)}|/|d_l^{(\beta)}|]$.
An experimentally accessible characteristic is the degree of linear polarization (DOP) of an exciton eigenstate, which is given by 
\begin{eqnarray}
\label{wzor:DOP}
P^{(\beta)}  & = & \frac{|d_l^{(\beta)}|^2-|d_t^{(\beta)}|^2}{|d_l^{(\beta)}|^2+|d_t^{(\beta)}|^2} \nonumber \\ 
 & = & -\frac{2}{\sqrt{3}} \frac{\im \left [ \alpha_{3/2,1/2}^{(\beta)*} 
\alpha_{-1/2,1/2}^{(\beta)}\right]}{  |\alpha_{3/2,1/2}^{(\beta)}|^2
+\frac{1}{3}|\alpha_{-1/2,1/2}^{(\beta)}|^2 }.
\end{eqnarray}
Clearly, the DOP is simply related to the angle $\chi^{(\beta)}$ by 
$P^{(\beta)}=\cos 2\chi^{(\beta)}$. Therefore, we will only present the results for the DOP.

For the simplified case of a system with Coulomb correlations omitted, the exciton states are identical to product states [$\beta \equiv i,j$] and the oscillator strengths used
in derivation of dipole moments are 
\begin{equation}\label{wzor:defalpha3}
\alpha_{3/2,1/2}^{(ij)} = \int d^3r \psi_{i}^{(\mathrm{h})}({\bm{r}})\psi_{j}^{(\mathrm{e})}({\bm{r}})\nonumber
\end{equation}
and
\begin{equation}\label{wzor:defalpha4}
\alpha_{-1/2,1/2}^{(ij)} =- \frac{1}{\Delta E_{\mathrm{lh}}} \int d^3r \psi_{i}^{(\mathrm{h})}({\bm{r}})R_k \psi_{j}^{(\mathrm{e})}({\bm{r}}).\nonumber
\end{equation}

For our numerical modelling, in the case of a simplified system without Coulomb correlations, we take into account several lowest lying states with dipole moments significant enough to give a contribution to optical response of the system. In the case of a Coulomb-correlated system, we restrict our computational basis to about 25 electronic states, and 70 hole states (exact values may vary depending on the structure). In the set of 70 hole states, we include also states with an excitation along the $y$ direction (n=2) which are coupled to n=0 states due to Coulomb interaction.

\section{Results}\label{s:results}
In this section, we present the results of our theoretical modelling of a single exciton confined in a QDash structure with an additional confining centre present. The change in the confinement strength is achieved by the change in the widening of the structure and described by the parameter $\lambda$. By adjusting the value of this parameter we show how the presence of the potential fluctuation leads to carrier trapping and study the consequences for optical absorption and emission. In order to find the minimal model needed to describe the qualitative changes introduced by the presence of the trapping centre we show how the system is modified by the introduction of Coulomb correlations.
First in section \ref{ss:sym}, we study a symmetrical system. Next, in section \ref{ss:asym}, we discuss the influence of geometrical asymmetry.
\begin{figure}[tb!]\label{fig:singlc}
\begin{center}
\includegraphics[height=84mm,angle=-90]{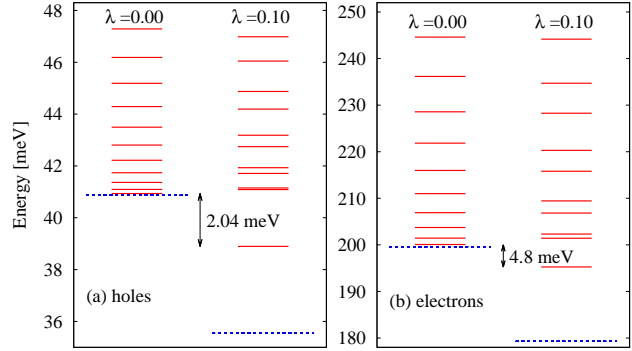}
\end{center}
\caption{\label{fig:single-carrier} Comparison of the energy spectrum of the single hole (a) and electron (b) eigenstates for a QDash without the widening (left hand side of each figure), and with 10\% widening present (right hand side of each figure). The bottom of the effective confinement potential, $\min[\epsilon_0(x)]$, has been shown in each case by a dashed line.}
\end{figure}

\subsection{Symmetrical widening}\label{ss:sym}
QDashes are relatively large structures hence one can expect small energy spacing between the lowest single carrier eigenstates (less than 1 meV). Introducing the widening of a structure changes this situation as the potential fluctuation traps the exciton ground state and shifts its energy down away from the excited states. In this paper, we will focus on the lowest energy eigenstates of the system since these states are of importance for the description of the structure in low and moderate temperatures and are expected to be strongly affected by the presence of the trapping centre. First, we present the data for a symmetrically placed QDash widening ($x_0=L/2$).

In figure~\ref{fig:single-carrier}, we show how the single carrier eigenenergies change after introduction of a 10\% widening of the central sector of a QDash (21\% increase in the volume of this sector). The ground state energy decreases by $2.04$ and $4.8$ meV for holes and electrons, respectively. The energy shift of higher states is much smaller, as they are not trapped by the potential fluctuation, thus the energy of only several lowest lying states will be strongly affected by the presence of the QDash widening. In figure~\ref{fig:single-carrier}, we have also presented the bottom of the effective potential (the bottom of the trapping fluctuation), as well as the bottom of the effective potential when there is no additional trapping present (or far away from the trapping centre). Although the trapping energy of an electron ground state is higher than that for a hole, the confinement of a hole is stronger, as it has a much higher effective mass and thus a lower probability density outside the trapping centre.
 
\begin{figure}[tb]
\begin{center}
\includegraphics[height=100mm,angle=-90]{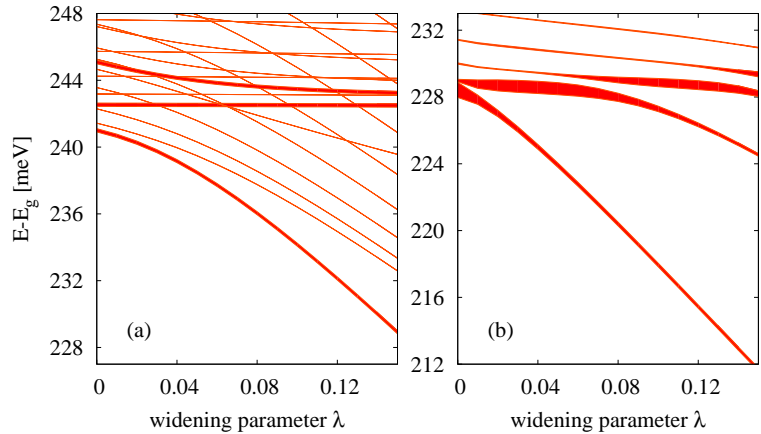}
\end{center}
\caption{\label{fig:maps2} The energy spectra for two cases: (a) without Coulomb interaction, (b) full model with Coulomb interaction included. The width of the lines is proportional to total absorption intensity $\sim |d_l|^2+|d_t|^2$. The scale of the line width is the same in both figures.}
\end{figure}

In order to assess the impact of Coulomb correlations we compare exciton spectra for two cases: with and without the Coulomb interaction, and study the spectra as a function of the widening of the QDash (figure~\ref{fig:maps2}). Apart from quite obvious shifts in the energy spectrum due to exciton formation, one can also see more subtle changes in the spectrum of the system. The energy spacings between the lowest energy eigenstates significantly vary with the change of the trapping centre strength. The observed decline in eigenenergies is a simple result of the change in single particle energies which decrease as the average potential felt by the carriers decreases. The strongest shift in energy is for the trapped state, which occupies the widened sector of the QDash. When the Coulomb interaction is included the energy spacing between the two lowest eigenstates highly increases as compared to the non-interacting case (from $3.7$ to $12.9$ meV for $\lambda=0.15$), which is a result of enhanced exciton binding energy and Coulomb-mediated interaction with higher energy excitonic states. 
Moreover, strong mixing between different excitonic states leads to a pronounced change in the absorption line intensities (which are proportional to the square of the total interband dipole moment). In the case of a system with no Coulomb correlations, the biggest contribution to optical response is from the exciton states with electron and hole envelope wave functions characterized by the same quantum numbers. In figure~\ref{fig:maps2}(a), only the three strongest states have this property. The exciton states with nearly constant energies consist of an electron and a hole in excited states. Since the electron wave function for the lowest excited state is odd, the hole wave function has also to be odd to yield nonzero overlap. For odd wave functions the carrier probability density in the centre of the structure is very small, thus the influence of the trapping potential is minimal. States with electrons and holes with different quantum numbers are bright as well, due to non-orthogonal electron and hole wave functions, but the oscillator strength of those states is usually more than ten times lower. The oscillator strength changes considerably after including the Coulomb interaction [figure~\ref{fig:maps2}(b)]. Due to mixing between different basis states, the oscillator strengths of previously nearly dark states, for certain values of QDash widening parameter, are much larger now. 
For certain values of the widening parameters, the oscillator strengths, and thus also absorption intensities of several lowest excited states, can be even higher than that of the ground state, which is quite the opposite to what is observed when the Coulomb interaction is neglected.
\begin{figure}[tb]
\begin{center}
\includegraphics[height=100mm,angle=-90]{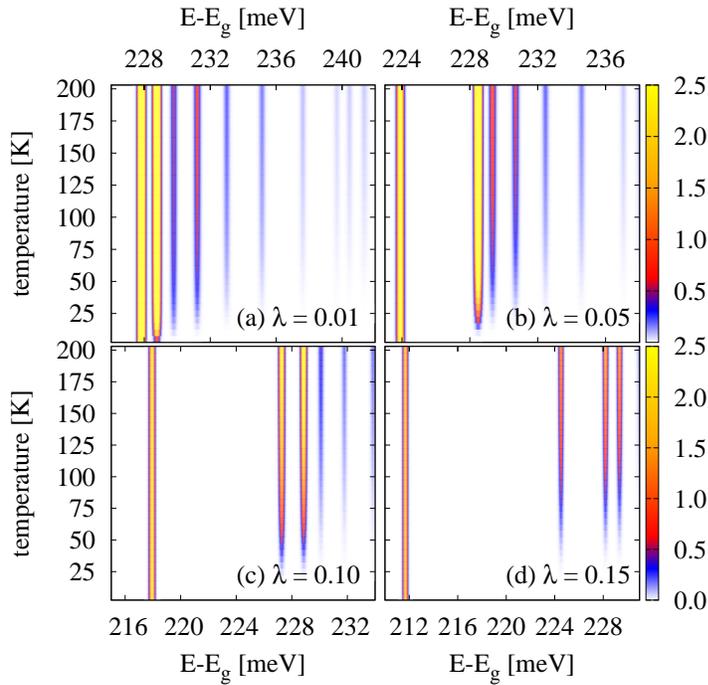}
\end{center}
\caption{\label{fig:maps-temp} The temperature dependent emission spectra for QDashes with four different values of a QDash widening. The emission lines are normalized to the intensity of the ground state. Boltzmann distribution of occupancies is assumed.}
\end{figure}
One can thus conclude that in the case of carrier trapping within an elongated structure, Coulomb interaction has a strong impact on the properties of the system and has to be taken into account. 

As we have seen, with increasing magnitude of the potential fluctuation the dipole moments of some excited states increase significantly. On the other hand, when modelling the emission intensities one has to take into account not only the dipole moments but also the distribution of occupations in the studied system. As the energy of the exciton ground state is strongly shifted down and the energy spacing between the ground and excited states increases with increasing depth of the trapping centre, the effect of enhancing the dipole moments of the excited states can be compensated in the optical emission from the structure by their reduced occupation. In figure~\ref{fig:maps-temp}, we show a temperature dependent map of the relative emission spectra for four different values of the widening parameter $\lambda$.  The Boltzmann distribution of occupancies of Coulomb correlated exciton states is assumed. As can be seen, the second emission line [figure~\ref{fig:maps-temp}(a,b)] is strongly visible even for relatively low temperatures (below 50~K) if the widening of the structure is small. A further increase of the value of the widening parameter [figure~\ref{fig:maps-temp}(c,d)] not only decreases slightly the oscillator strength of this line but also, due to a shift in energy, decreases the occupancy of this state. Therefore, the line is visible only at higher temperatures. Similar effects of decline in occupancy is observed for higher states, and if not for strong enhancement of their transition rates, intensity of those lines would be very small.

\begin{figure}[tb]
\begin{center}
\includegraphics[width=100mm]{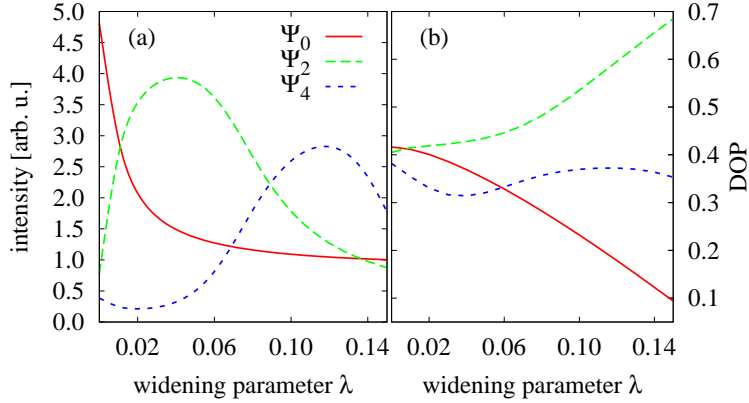}
\end{center}
\caption{\label{fig:trdop} The dependence of the total absorption intensity (a) and the degree of linear polarization (b) of several lowest lying excitonic states on the widening of the QDash. The omitted exciton eigenstates $\Psi_1$ and $\Psi_3$ are composed of single particle envelope wave functions of opposite parity and therefore are dark.}
\end{figure}

In figure~\ref{fig:trdop}, we quantitatively describe the change in the absorption intensities for several lowest energy excitonic lines for the two orthogonal polarizations of the light. In order to characterize both polarizations it is sufficient to present the total intensity of a line, $|d_l|^2+|d_t|^2$, and its degree of linear polarization (\ref{wzor:DOP}).
The following study explains how certain states govern the $S$--shaped DOP temperature dependence observed in the experiment \cite{musial12} and how it would be influenced by the changes in the strength of the confining potential.

In figure~\ref{fig:trdop}(a), we present the dependence of the absorption intensity of several lowest energy eigenstates on the magnitude of the confining potential fluctuation. The characteristic feature of these intensities is the decline in the intensity of the ground state and a non-monotonic behaviour of excited states. The decline in the intensity of a the ground state can be attributed to a small decline in the electron and hole overlap for higher magnitudes of the trapping centre (less than 10\% of the observed decline), as well as to strong changes in the values of the Coulomb matrix elements, and therefore changes in the composition of the system eigenstates. In this figure, the enhancement of the intensities of excited states is clearly visible. The ranges of the $\lambda$ parameter for which the intensity of the excited state is larger than that of the ground state is quite wide. Furthermore, the intensities for those states at their maximum are up to 10 times higher than in the absence of the trapping centre.

In figure~\ref{fig:trdop}(b), we present the degree of linear polarization for three lowest exciton bright states as a function of the value of the widening parameter $\lambda$. With the increasing strength of the trapping potential the characteristic confinement length of the exciton ground state in the elongation direction decreases. If one refers to a simplified relation between the geometry of the structure and the DOP, derived for a uniform system without Coulomb interaction \cite{kaczmarkiewicz11a},
\begin{equation}
P \sim \frac{n_t^2}{D^2} - \frac{n_l^2}{L_\mathrm{eff}^2},\nonumber
\end{equation}
where $D$ and $L_\mathrm{eff}$ are the width and the effective confinement length of the structure, respectively, and $n_{t(l)}$ is the quantum number describing excitation in the transverse (elongation) direction, one immediately sees that the more isotropic the wave function is, the lower degree of linear polarization can be observed. When the confining potential gets stronger the confinement length in the elongation direction gets closer to the confinement length in the transverse direction, thus increasing the overall ground state isotropy, leading to a more isotropic emission from the exciton ground state.
The DOP of the excited states evolves in a nontrivial way with increasing the strength of the confining potential. However, the results shown in figure~\ref{fig:trdop} allow one to predict that thermal filling of higher excitonic states which show higher DOP will lead to an initial increase of the overall DOP from the structure. At still higher temperature, occupation of even higher states will become considerable. As these states tend to have weaker polarizatlon properties at moderate values of $\lambda$ the overall DOP will saturate or even decrease, leading to an $S$-shaped temperature dependence of the DOP.

\begin{figure}[tb]
\begin{center}
\includegraphics[width=100mm]{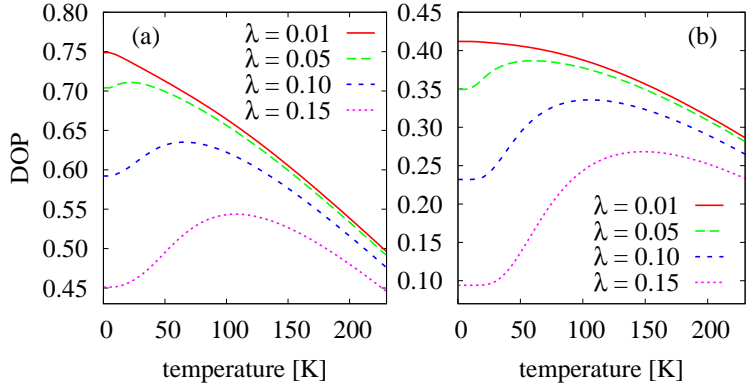}
\end{center}
\caption{\label{fig:dopc_nc} The DOP temperature dependence for different values of widening parameter $\lambda$ (a) Coulomb interaction omitted (b) full model. In calculations Boltzmann distribution of occupancies is assumed. The figures represent integrated degree of polarization from all occupied states.}
\end{figure}

In figure~\ref{fig:dopc_nc}, we present the temperature dependence of the DOP for several values of the widening parameter for the simplified case without Coulomb interaction and for the full model described by \Eref{wzor:ham_ex}. The presented DOP is calculated by adding up the emission from all the occupied states in a single QDash (assuming thermal distribution of occupancies). As can be seen, even for relatively low values of the widening parameter, the characteristic $S$-shaped curve can be observed. Although the results for both cases are qualitatively similar, the quantitative differences are rather large. First of all, the overall DOP values for the simplified case [figure~\ref{fig:dopc_nc}(a)] are much larger than that calculated for the full model [figure~\ref{fig:dopc_nc}(b)]. Also, the difference between the degree of polarization at low temperatures and its maximum at moderate temperatures is smaller when the Coulomb interaction is neglected. This contradicts the experiment \cite{musial12}, where the observed changes in the DOP were even larger than those presented in figure~\ref{fig:dopc_nc}(b). Even though neglecting the Coulomb interaction can be useful for grasping overall tendencies in the behaviour of the system, the quantitative discrepancies preclude using such a simplified model in a more detailed analysis.

\subsection{Asymmetrical QDash}\label{ss:asym}
As the QDash structures are often highly irregular \cite{reithmaier07} the case of a symmetrically placed localizing centre corresponds to a rather idealized situation,
in this section we address the influence of the position of the trapping centre on the electronic and optical properties of a QDash. The shift from the QDash centre is described by the parameter $x_0$. In the case of a symmetrically placed trapping centre the spectrum can be decomposed into two parts: one optically active, and the other one optically inactive due to selection rules (with excitonic states constructed as a products of single particle states with $x$-dependent wave function components of opposite parity). In the case of a symmetrical QDash, there are no Coulomb interaction matrix elements leading to mixing between those two groups of states. When the geometry of the QDash is changed by a shift of the trapping centre the states from both groups will have non-zero oscillator strengths and, moreover, new Coulomb interaction matrix elements will appear leading to spectrum reconstruction. Since in the limiting case of a symmetrical QDash the spectrum should again split into two decoupled groups of states one can expect relatively small changes in the spectra for small shifts of the widening. 

\begin{figure}[tb]
\begin{center}
\includegraphics[height=100mm,angle=-90]{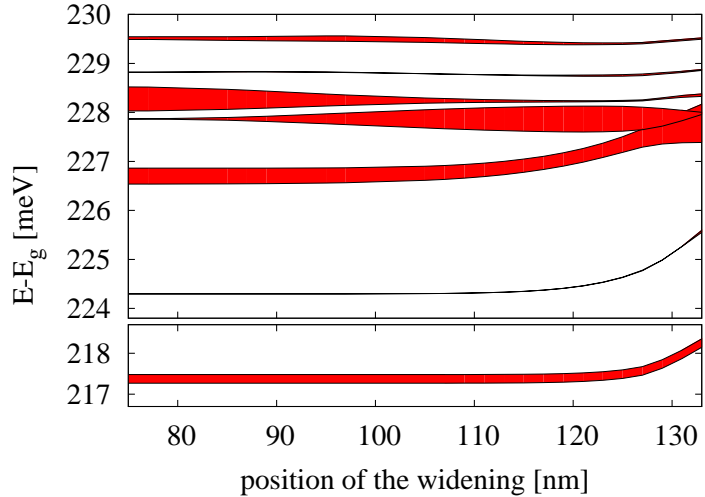}
\end{center}
\caption{\label{fig:mapsasym} The QDash energy spectrum dependence on the position of the widening. Several lowest exciton states have been shown. The width of the lines is proportional to the total absorption intensity $\sim |d_l|^2+|d_t|^2$.}
\end{figure}

In figure~\ref{fig:mapsasym}, we show the energy spectrum of a QDash as a function of the position of the trapping centre, for the widening parameter $\lambda=0.1$. The width of the lines is proportional to the absorption intensity. Only a small shift of the exciton eigenenergies is noted as the trapping centre shifts toward the edge of the structure (less than 1 meV for the strongest shift and less than 0.1 meV for moderate shifts). Again, the absorption intensity of higher excited states changes strongly, and can be larger than that of the ground state.
\begin{figure}[tb]
\begin{center}
\includegraphics[width=100mm]{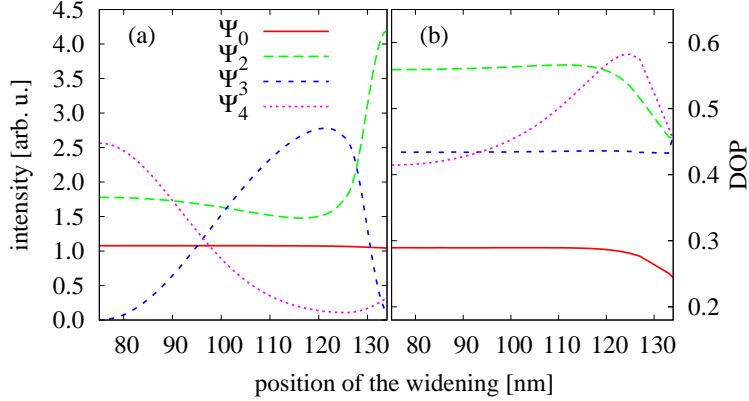}
\end{center}
\caption{\label{fig:asymdop} The dependence of the total absorption intensity (a) and the degree of linear polarization (b) of several lowest lying excitonic states on the position of the QDash widening ($\lambda=0.1$). Only lowest energy states, with significant values of transition rate have been presented.}
\end{figure}

In figure~\ref{fig:asymdop}, we present the total absorption intensity and the degree of linear polarization as a function of the position of the widening of the QDash. The intensity of the ground state shows very little change with the change of the trapping centre location. On the other hand there is quite a large enhancement of the intensity of the $\Psi_2$ exciton state, but only for strong shifts of the trapping centre. As the position of the trapping centre is expected to be random in an ensemble of QDashes, the average intensity of this state will be much smaller than that, closer to that of an idealized symmetrical QDash. Also, the emission from this state is more isotropic for strong shifts of the trapping centre, which leads to a decline in the DOP. There is a very strong change in the intensities of the two highest exciton states shown in this figure. The state $\Psi_3$ is dark for a symmetrical QDash but its intensity increases significantly when the trapping centre shifts. The $\Psi_4$ state behaves in the opposite manner: for the symmetrical case its intensity is relatively large and decreases when the widening shift is introduced. These two states will have a similar effect on the overall degree of linear polarization. When the intensity of any of those two states is significant the DOP is moderate, between 0.4 and 0.45. Sudden increase of the DOP of $\Psi_4$ state will have very little effect on the properties of the system, as for large shifts of the trapping centre its intensity is very small. By comparing intensities with the DOP properties one can assess the temperature dependence of the latter. Similarly to the symmetrical case, the highly trapped state has the lowest DOP, the DOP of the first excited state is significantly higher and the two highest presented excited exciton states have moderate values of the DOP. Upon thermal filling of the states, one can again explain the S-shaped temperature dependence of the DOP observed in the experiment as the polarization properties are not qualitatively different from the symmetrical QDash system [figure~\ref{fig:dopc_nc}(b)]. Moreover, shifts in the position of the widening of a structure are of lesser importance in modelling the system properties, as they do not change its overall qualitative properties and even the quantitative changes are relatively small.

\section{Conclusions}\label{s:conclusions}
We have investigated the influence of the presence of an additional trapping centre 
and its geometry on the electronic structure and optical properties 
of elongated self-assembled nanostructures (quantum dashes). We focused on the magnitude of the confining
potential fluctuation and on its position. We have shown
that a strong shift in the position of the trapping centre
leads to the appearance of additional bright states and also
changes the intensities of other optically active
states. We show also how Coulomb correlations influence the properties  
of the structure  and study if they are of importance in 
reproducing the characteristic features of the system.
The presented results confirmed that proper modelling of an 
exciton confined in the structure must account for Coulomb 
interaction, as it leads to large changes in exciton spectra.
Also, the presence of the trapping potential 
fluctuation leads to qualitative and quantitative changes 
in the system spectrum and needs to be accounted for
when modelling such highly irregular structures. 
Although the exact position of the trapping centre
influences the energy spectrum and dipole moments, 
it does not considerably affect the polarization properties of the system.

Our detailed study confirms previous predictions concerning 
polarization dependence of radiation emitted by QDashes. 
The degree of polarization of the ground state emission decreases as
the strength of the trapping centre increases and the evolution of
the degree of polarization of higher excited states leads
to an $S$-shaped DOP temperature dependence.
We have also shown that the properties of an individual QDash
strongly depend on the details of the confining potential. Strong 
enhancements of the line intensities can be observed, as well as strong 
reductions, depending on the confinement properties. 

Our results may be useful for the design of nanostructure-based devices, in particular single-photon emitters, where excitonic effects, modulation of light intensities and detailed polarization properties are of practical importance.

\ack
This work has been supported in parts by the Polish Ministry of
Science and Higher Education (Grant No. N N202
181238) and by the TEAM programme of the 
Foundation for Polish Science, co-financed by the European Regional
Development Fund. PK acknowledges support from German Academic
Exchange Service (DAAD). The authors would like to thank A. Musia{\l} and G. S\k{e}k
for interesting discussions and for many useful comments and suggestions.

\section*{References}
\bibliographystyle{unsrt}

\end{document}